\documentclass[aps,prb,twocolumn,floatfix,notitlepage, superscriptaddress]{revtex4-2}
\usepackage[utf8]{inputenc}
\usepackage{amsfonts,amsmath,amssymb,graphicx,epstopdf}
\usepackage{xcolor}
\usepackage{natbib}
\usepackage[colorlinks]{hyperref}
\definecolor{darkred}{rgb}{0.5,0,0}
\definecolor{darkgreen}{rgb}{0,0.5,0}
\definecolor{darkblue}{rgb}{0,0,0.5}
\hypersetup{colorlinks,
linkcolor=darkred,
filecolor=darkgreen,
urlcolor=darkblue,
citecolor=darkgreen}
\usepackage{braket}
\usepackage{overpic}

\newcommand{\nep}{\operatorname{e}}

\usepackage{orcidlink}

\begin{document}
\title{Logarithmically slow heat propagation in a clean Josephson-junction chain}

\author{Angelo Russomanno~\orcidlink{0009-0000-1923-370X}}
\affiliation{Dipartimento di Fisica ``E. Pancini'', Universit\`a di Napoli Federico II, Complesso di Monte S. Angelo, via Cinthia, I-80126 Napoli, Italy}
\begin{abstract}
  We consider a clean Josephson-junction chain coupled by one of its extremities to a thermal bath through a resistance. Considering the Langevin dynamics in the classical regime, in the case of Josephson energy much smaller than charging energy, we find that heat propagates logarithmically slowly through the system, rather than diffusively, as highlighted by the logarithmic increase in time of a thermalization length we define and by the logarithmically slow increase in time of the energy. This behavior -- typical of quantum Anderson or many-body localized systems -- is observed here also in a clean classical glassy Hamiltonian system. We argue that this phenomenon might imply strong robustness to the effect of ergodic inclusions for the nonergodic behavior in the charge-quantized regime.
\end{abstract}
\maketitle
Junctions between two superconductors separated by a normal region exhibit rich physics. These so-called superconductor-normal-superconductor (SNS) Josephson
junctions (JJs) allow dissipationless current to flow through the normal region, whose direction is determined by the phase difference between the two superconductors~\cite{Tinkham}.
In the last decades, this fundamental unit has been extended to arrays of JJs~\cite{FAZIO2001235,geige}. Among the myriad of phenomena observed in JJ arrays are the superconductor-insulator transition~\cite{goldman,goldbart,bottcher}, achieved by varying the ratio between the Josephson coupling between islands to the charging energy of the islands, and the Berezinskii-Kosterlitz-Thouless transition driven by proliferation of vortices which turn the superconductor to a normal metal~\cite{Jos201340YO,PhysRevB.110.L180502}. They can also display chiral topological phases~\cite{PhysRevB.109.144519}, a topological superinsulator phase~\cite{condmat8040097}, and Thouless topological pumping~\cite{10.21468/SciPostPhys.16.3.083}. They can be moreover used to simulate many model Hamiltonians, from the clock and the tricritical Ising models~\cite{PhysRevB.111.045418,PhysRevLett.132.226502}, to the frustrated XY model~\cite{Ciria_1999} and the continuous and discrete sine-Gordon systems~\cite{VANDERZANT1998219,Mazo2014}. Josephson-junction arrays can also display persistent time-translation symmetry breaking oscillations~\cite{2yzx-jfky,kxkc-4gs4,PhysRevB.108.094305} similar to time crystals (for a review on time crystals see~\cite{RevModPhys.95.031001,khemani2019briefhistorytimecrystals}).

Another interesting aspect of Josephson junction arrays is the possibility they have to display ergodicity breaking effects, both in the classical and quantized-charge regimes, similar to the many-body localization (MBL -- see~\cite{RevModPhys.91.021001} for a review). It has been argued that at high temperatures and small Josephson couplings the energy cost of a unit charge transfer between two sites exceeds the matrix element of the charge transfer, leading to a MBL-like behavior in a Josephson-junction chain~\cite{pino}. Beyond that, in a strictly related model -- a strongly-interacting Bose-Hubbard lattice -- an extremely slow thermalization~\cite{carleo} and a nonergodic behavior in the small-hopping regime~\cite{Russomanno_2020,PhysRevA.90.033606,PhysRevLett.105.250401} have been found, and the same Josephson-junction chain studied in~\cite{pino} displays at intermediate energies a weak ergodicity breaking akin to quantum scars~\cite{Russomanno_2022}. Beyond that, this same Josephson-junction chain in the classical regime has been observed to display a glassy behavior with thermalization time many order of magnitudes larger than the inverse largest Lyapunov exponent~\cite{Mithun_2019}.

In this context of ergodicity breaking, an interesting phenomenon to study is the propagation of heat through a one-dimensional system, once one of its extremities has been coupled to a thermal bath. This approach has been used in MBL systems to study the slowest thermalization time, and see if it scales faster than the Heisenberg time, in order to assess robustness to ergodic inclusions~\cite{PhysRevB.106.L020202,PhysRevB.105.174205,PhysRevB.107.014203,shen2026quantumavalanchestabilitymanybody}. In this framework, both in Anderson and many-body localized systems, a peculiar logarithmic propagation front of the heat has been observed~\cite{PhysRevB.110.134204,L_M_Lezama_2022,bhakuni2023noiseinducedtransportaubryandreharpermodel}. This logarithmic thermalization front, so different from the usual diffusive one, gives rise to a slowest thermalization time exponential in the system size -- making a scaling faster then the Heisenberg time possible -- and is consistent with the usual logarithmic lightcones in MBL systems~\cite{kim2014localintegralsmotionlogarithmic,Elgart_2024,toniolo2024stabilityslowhamiltoniandynamics}.

In this letter we apply a similar analysis to a clean Josephson-junction array in the classical regime (the one without charge quantization~\cite{pino}). We couple it by one of its extremities to a thermal bath and we numerically describe it using a Langevin-equation approach. In the regime of Josephson energy much smaller than the charging energy we find a logarithmic propagation front of the heat, as witnessed by the logarithmic increase in time of a thermalization length analogous to the one defined in~\cite{PhysRevB.110.134204}. This very slow heat propagation, so different from the diffusive behavior one would naively expect, is consistent with the extremely slow thermalization times observed in~\cite{Mithun_2019}. It is strictly similar to the one observed in MBL systems and provides a different perspective on the glassy behavior of this model described in~\cite{Mithun_2019}. The model is a classical glassy clean one and displays a similar behavior of the energy propagation in quantum disordered systems. To the best of our knowledge, this is the frist time such a phenomenon is observed. 

Together with the slow propagation of heat we find also a slow energy increase: the energy increases logarithmically in time and the increase starts after a long prethermal plateau. So we find here a classical prethermalization phenomenon due to the coupling of a Hamiltonian system to a thermal bath (and the related Gaussian Langevin noise). This provides a different physical mechanism for classical prethermalization that is usually achieved in periodically driven systems~\cite{PhysRevLett.127.140602,HO2023169297,Howell2019,PhysRevB.100.100302,Ye_2021}. The logarithmic increase of the energy and the one of the thermalization length can be related through a simple analytcal model.
\begin{figure}
  \centering
  \vspace{0.5cm}
  \includegraphics[width=80mm]{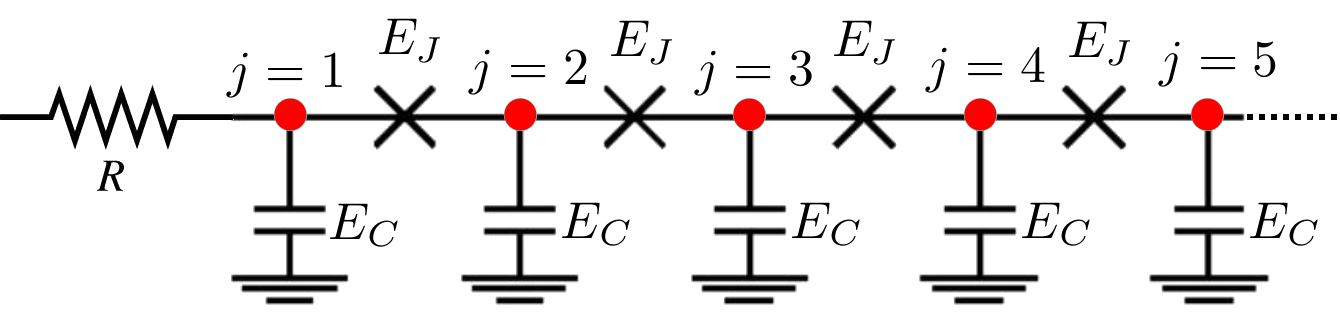}
  \caption{Circuit scheme of the Josephson junction chain. Notice the coupling to the resistance through the leftmost site. Each node (red circles) corresponds to a superconducting island and is labeled with an index $j$ going from 1 to $L$. Each capacitor has a capacitance $C$ and a charging energy $E_C = (2e)^2/C$, and each Josephson junction a Josephson energy $E_J$.}\label{circ:fig}
\end{figure}

The circuit scheme is the one provided in Fig.~\ref{circ:fig}. Assuming $\hbar = 1$, the Langevin dynamics is provided by the equations
\begin{align}\label{dyn:eqn}
  \dot{q}_j &= -\partial_{\theta_j}\mathcal{H}(\{\theta_j,q_j\}) -\frac{1}{RC} q_1\delta_{1j} + \delta_{1j}\xi(t)\,,\nonumber\\
  \dot{\theta}_j&=\partial_{q_j} \mathcal{H}(\{\theta_j,q_j\})\,,
\end{align}
where $q_j$ is the charge on the $j$-th superconducting island and $\theta_j$ the corresponding superconducting phase. The Hamiltonian is given by
%
  \begin{equation}\label{Ham:eqn}
    \mathcal{H}(\{\theta_j,q_j\}) = \frac{E_C}{2} \sum_{j=1}^L q_j^2 -E_J\sum_{j=1}^{L-1}\cos(\theta_j-\theta_{j+1})\,.
  \end{equation}
%
In the equations above $C$ is the value of the capacitances, $E_C= (2e)^2/C$ is the charging energy, $E_J$ the Josephson energy, and $R$ the value of the resistance. Notice that the resistance (that's to say the coupling to the thermal bath) acts only on the first site, as highlighted by the Kronecker delta symbol $\delta_{1j}$. Also the thermal noise associated to the resistance acts only on the first site and is a Gaussian process with vanishing average ($\braket{\xi(t)}=0$) and variance given by $$\braket{\xi(t)\xi(t')}=\frac{2 k_B T}{RCE_C}\delta(t-t')\,,$$ where $k_B$ is the Boltzmann constant. The variance has been chosen so that in the absence of Josephson coupling -- that's to say for $E_J = 0$ -- the Fokker-Planck equation provides a Boltzmann steady-state distribution of the form $p\propto\nep^{- E_C q_1^2 / (2 k_B T)}$. This Fokker-Planck analysis has been performed in the standard way explained for instance in~\cite{Hannes,Frey}. Eq.~\eqref{dyn:eqn} has been obtained simply applying the Kirchhoff laws to the circuit shown in Fig.~\ref{circ:fig}, as explained in~\cite{devo} (see an example of this analysis in~\cite{PhysRevB.108.094305}). In what follows we are going to assume $E_C = 1$.

We perform the numerics using the Verlet algorithm with noise. We divide the time in discrete steps of length $\Delta t$ and define $q_j^{(n)} \equiv q_j(n\Delta t)$ and $\theta_j^{(n)} \equiv \theta_j(n\Delta t)$. The evolution step is given by
\begin{align}
  q_j^{(n+1)}&  = q_j^{(n)} -\partial_{\theta_j}\mathcal{H}(\{\theta_j^{(n)},q_j^{(n)}\})\Delta t -\frac{q_1^{(n)}}{RC} \delta_{1j}\Delta t + \delta_{1j}W_n\nonumber\\
  \theta_j^{(n+1)}&  = \theta_j^{(n)} + \partial_{q_j} \mathcal{H}(\{\theta_j^{(n)},q_j^{(n+1)}\}) \Delta t\,,
\end{align}
where $W_n$ are uncorrelated Gaussian random variables with variance $\sigma^2(W_n) = 2 \tau \Delta t$, where we have defined the reduced temperature as $\tau \equiv k_B T/(RCE_C) = k_B T/(4 e^2 R)$. We consider average over $N_{\rm r}$ realizations of the randomness process and we have observed convergence for $\Delta t = 10^{-4}$ and $N_{\rm r} = 990$. The error on the average over the number of realizations is provided by the root mean square fluctuation divided by $\sqrt{N_{\rm r}}$. In all the simulations we initialize the system in a state of vanishing charges $q_j$ and $\theta_j$ randomly taken from a uniform distribution in the interval $[0,2\pi]$.

\begin{figure}
  \centering
  \begin{tabular}{c}
    (a)\\
    \includegraphics[width=80mm]{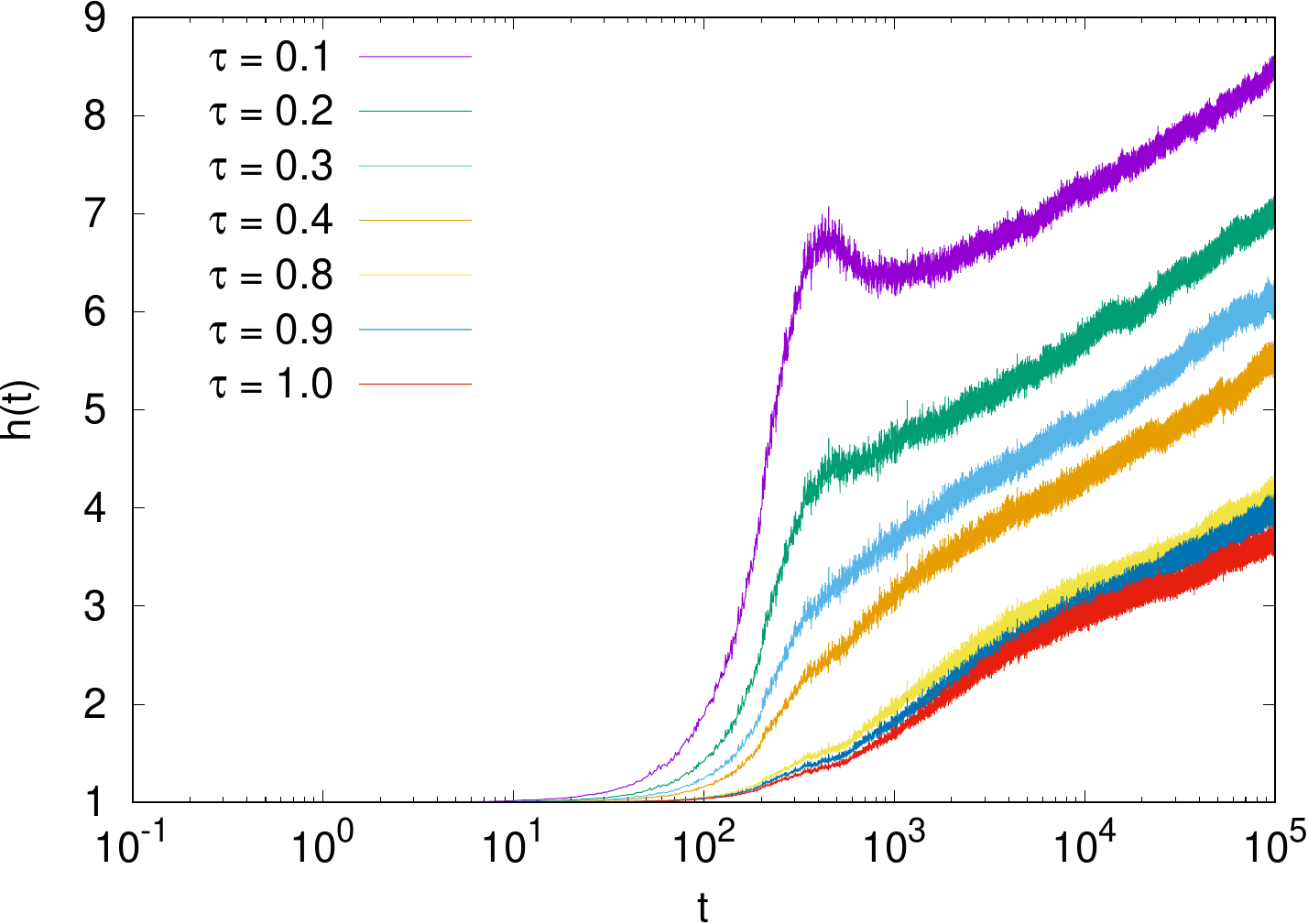}\\
    (b)\\
    \includegraphics[width=80mm]{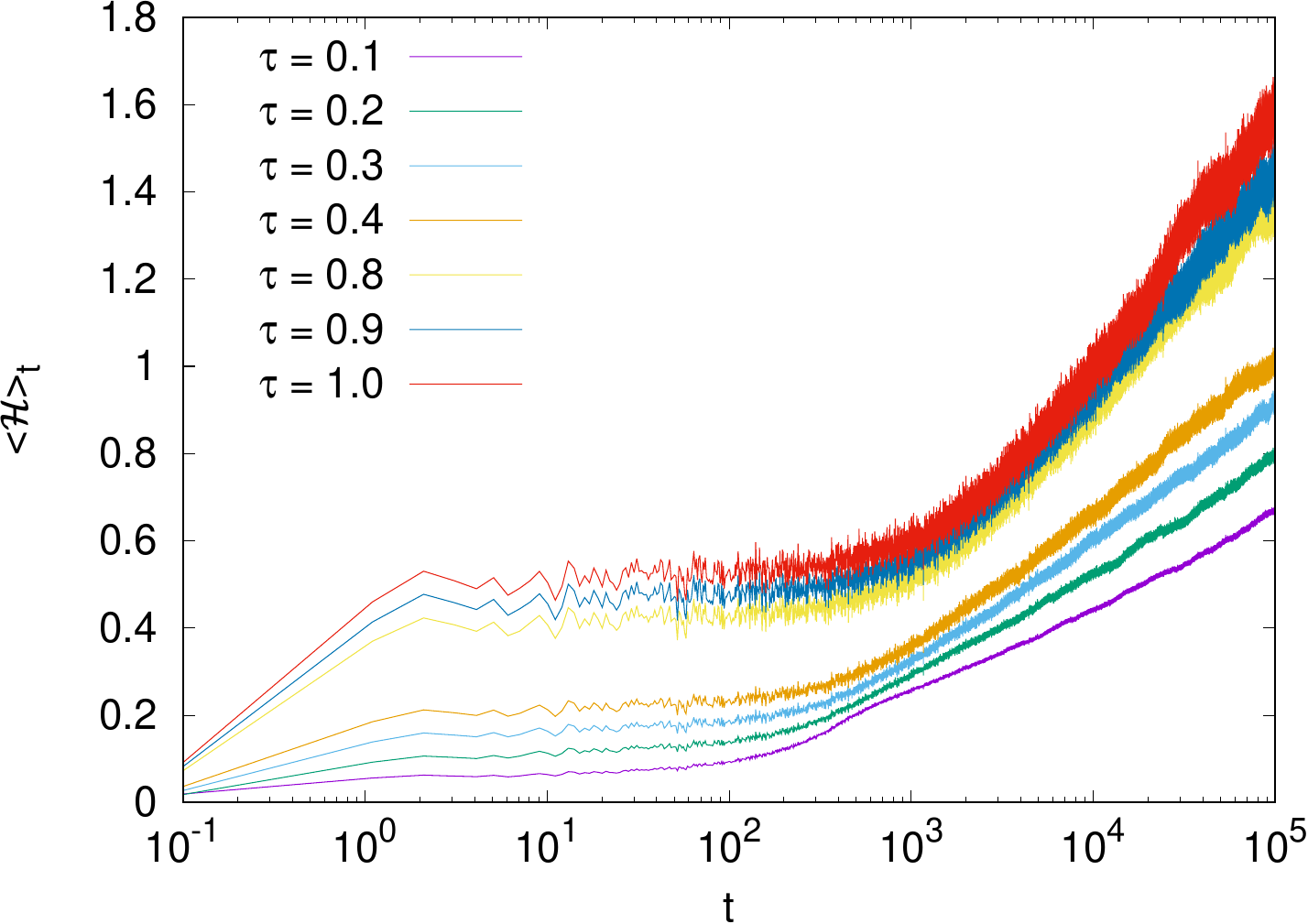}
  \end{tabular}
  \caption{(Panel a) Thermalization length $h(t)$ [see Eq.~\eqref{leng:eqn}] versus time $t$, with the horizontal axis plotted in a logarithmic scale, for different values of the reduced temperature $\tau$. We see a quite  clear logarithmic increase after a prethermal transient where $h(t)$ is vanishing. (Panel b) Energy $\braket{\mathcal{H}}_t$ versus time $t$ with the horizontal axis in a logarithmic scale. Notice the logarithmic increase in time after an initial prethermal plateau, quite evident for the larger values of the reduced temperature $\tau$. Numerical parameters: $N_{\rm r} = 990,\,\Delta t = 10^{-4},\,E_J = 10^{-2},\,RC = 1,\,E_C = 1,\,L=20$.}\label{leng:fig}
\end{figure}

We consider the case $E_J \ll E_C$ and we define a thermalization length in the following way
\begin{equation}\label{leng:eqn}
  h(t) = \frac{\sum_{j=1}^Lj\braket{q_j^2}_t}{\sum_{j=1}^L\braket{q_j^2}_t}\,,
\end{equation}
where $\braket{\ldots}_t$ marks the average over randomness realizations at time $t$. This definition of thermalization length is in analogy with a similar definition in~\cite{PhysRevB.110.134204} and probes at time $t$ the rightmost site where the charging energy has reached is thermal equilibrium value. We plot $h(t)$ versus time $t$ for $E_J=10^{-2}E_C$, $L=20$ and different values of the rescaled temperature $\tau$ in Fig.~\ref{leng:fig}(a). The logarithmic increase in time is unmistakable (notice the logarithmic scale on the horizontal axis) and, quite remarkably, the increase is faster for smaller values of the temperature.

The same logarithmic increase in time can be observed in the energy Eq.~\eqref{Ham:eqn} averaged over the Langevin trajectory ensemble. For larger values of $\tau$ the increase starts after a long-lived prethermal plateau. We can understand the logarithmic increase of the energy from the one of the thermalization length. Being $E_J \ll E_C$ we can approximate the energy as the charging energy. With a very rough approximation we can say that at time $t$ all the sites with $j < h(t)$ have attained the thermal value of the charging energy $\braket{E_C q_j^2 / 2} \simeq k_B T/2$ (neglecting the Josephson energy one can apply the equipartition theorem). So, in this approximation, the energy at time $t$ is given by
\begin{equation}\label{eqi:eqn}
  \braket{\mathcal{H}}_t \simeq \frac{k_B T}{2} h(t)\,,
\end{equation}
and so if $h(t)$ displays a logarithmic increase the same is done by $\braket{\mathcal{H}}_t$. Because in Eq.~\eqref{eqi:eqn} $h(t)$ is multiplied by $T$, one can also understand how smaller slopes for larger temperature in Fig.~\ref{leng:fig}(a) can provide the opposite in Fig.~\ref{leng:fig}(b).

So, in summary, we have found in a clean classical glassy system a logarithmic light cone of the heat, a phenomenon that has before observed only in disordered localized quantum systems. The finding is interesting because here -- in contrast with quantum localized systems -- there are no localized integrals of motion, so the physical mechanism leading to this phenomenon is different. One possibility of future research would be to consider the case of a quantized Josephson-junction chain. There might be the possibility that a logarithmically slow increase in energy comes together with a logarithmically slow increase in entanglement entropy (as occurs in MBL systems), and so an efficient matrix-product state numerical analysis would be possible. 

If also in the quantized case a logarithmic propagation front of the heat were found, this would have interesting implications for the robustness of nonergodicity when coupling the system to an ergodic inclusion. When $E_J \ll E_C$ the spectrum is organized in narrow bands separated by gaps~\cite{pino}. The $n$-th band that lies around energy $E_n = E_C n^2 / 2$ is separated by the next one by a gap $\Delta_n = E_C (2n+1) / 2$. Inside each band there are $L$ states, so the gaps {\em inside} a band scale like $1/L$, and therefore the Heisenberg time (that scales like the inverse of this gap) is linear in $L$. With the logarithmically slow propagation of heat, the Thouless time -- that's to say the slowest thermalization time -- estimated as the time needed for the thermalization front to fill the whole chain, scales exponentially in $L$ (see the analysis in~\cite{PhysRevB.110.134204}). So the Thouless time is expected to scale much faster than the Heisenberg time, and so -- applying the argument of~\cite{PhysRevB.105.174205,PhysRevB.106.L020202} -- we see that nonergodicity in this model in the quantum regime is expected to be very robust to ergodic inclusions. 
\acknowledgements{We acknowledge financial support from PNRR MUR Project PE0000023-NQSTI.}
%
%
%
\end{document}